\documentclass[aps,prd,showpacs,preprintnumbers]{revtex4}
\usepackage{amsmath,amssymb}
\usepackage{graphicx}
\usepackage{subfigure}

\newcommand\bef{\begin{figure}}
\newcommand\eef[1]{\label{fg:#1}\end{figure}}
\newcommand\beq{\begin{equation}}
\newcommand\eeq[1]{\label{#1}\end{equation}}
\newcommand\beqa{\begin{eqnarray}}
\newcommand\eeqa[1]{\label{#1}\end{eqnarray}}
\newcommand\bet{\begin{table}}
\newcommand\eet[1]{\label{tb:#1}\end{table}}
\newcommand\betb{\begin{center}\begin{tabular}}
\newcommand\eetb{\end{tabular}\end{center}}
\newcommand\beit{\begin{itemize}}
\newcommand\eeit{\end{itemize}}

\newcommand\fgn[1]{Figure \ref{fg:#1}}
\newcommand\eqn[1]{eq.\ (\ref{#1})}
\newcommand\scn[1]{Section \ref{s:#1}}
\newcommand\tbn[1]{Table \ref{tb:#1}}
\newcommand\apxn[1]{Appendix \ref{s:#1}}

\newcommand\incfig[2]{\includegraphics[scale=#1]{#2}}

\newcommand\AV{{\scriptscriptstyle AV}}
\newcommand\eint{E_{\rm int}}
\newcommand\had{{\scriptscriptstyle H}}
\newcommand\N{{\scriptscriptstyle N}}
\newcommand\Nm{{\scriptscriptstyle N_-}}
\newcommand\Np{{\scriptscriptstyle N_+}}
\newcommand\PS{{\scriptscriptstyle PS}}
\newcommand\SC{{\scriptscriptstyle S}}
\newcommand\SW{{\scriptscriptstyle SW}}
\newcommand\V{{\scriptscriptstyle V}}
\newcommand\x{{\mathbf x}}

\newcommand\ie{{\sl i.e.\/}}

\newcommand\etal{{\sl et al.\/}}

\newcommand\epj{{\sl Eur.\ Phys.\ J.\/}\ }
\newcommand\jhep{{\sl J.\ H.\ E.\ P.\/}\ }
\newcommand\np{{\sl Nucl.\ Phys.\/}\ }
\newcommand\npbps{{\sl Nucl.\ Phys.\/} B ({\sl Proc. \ Suppl.\/})\ }
\newcommand\pos{{\sl PoS\/}\ }
\newcommand\pr{{\sl Phys.\ Rev.\/}\ }
\newcommand\pls{{\sl Phys.\ Lett.\/}\ }
\newcommand\prlt{{\sl Phys.\ Rev.\ Lett.\/}\ }

\begin{document}
\title{Nucleons near the QCD deconfinement transition}
\author{Saumen\ \surname{Datta}}
\email{saumen@theory.tifr.res.in}
\affiliation{Department of Theoretical Physics, Tata Institute of Fundamental
         Research,\\ Homi Bhabha Road, Mumbai 400005, India.}
\author{Sourendu\ \surname{Gupta}}
\email{sgupta@theory.tifr.res.in}
\affiliation{Department of Theoretical Physics, Tata Institute of Fundamental
         Research,\\ Homi Bhabha Road, Mumbai 400005, India.}
\author{M.\ \surname{Padmanath}}
\email{padmanath@theory.tifr.res.in}
\affiliation{Department of Theoretical Physics, Tata Institute of Fundamental
         Research,\\ Homi Bhabha Road, Mumbai 400005, India.}
\author{Jyotirmoy\ \surname{Maiti}}
\email{jyotirmoy.maiti@gmail.com}
\affiliation{Department of Physics, Barasat Government College, Barasat,
         Kolkata 700124, India.}
\author{Nilmani\ \surname{Mathur}}
\email{nilmani@theory.tifr.res.in}
\affiliation{Department of Theoretical Physics, Tata Institute of Fundamental
         Research,\\ Homi Bhabha Road, Mumbai 400005, India.}
\begin{abstract}
Using non-perturbative lattice method we studied hadronic screening
correlators above and immediately below the deconfinement transition
temperature, $T_c$, in the quenched approximation with lattice spacing of
$1/(8T)$ using clover improved Wilson fermions. Simulations were performed
at temperatures $T/T_c = 0$, $0.95$ and $1.5$.  Mesonic screening
correlators show no statistically significant thermal effects below
$T_c$, and clear evidence for weakly interacting quarks above $T_c$.
Baryon screening correlators yield similar physics above $T_c$, but show
precursor effects for chiral symmetry restoration below $T_c$.
\end{abstract}
\pacs{12.38.Mh, 11.15.Ha, 12.38.Gc}
\preprint{TIFR/TH/12-48}

\maketitle

\section{Introduction\label{s:Intro}}

\bet[b]
\begin{center}\begin{tabular}{|c||c|c||c|c|}
\hline
$\frac{m_\pi}{T_c}$ & 
$\frac{\mu_\V}{m_\V}$ & $\frac{\mu_\Np}{m_\Np}$ &
$\frac{\mu_\AV-\mu_\V}{m_\AV-m_\V}$ & 
$\frac{\mu_\Nm-\mu_\Np}{m_\Nm-m_\Np}$ \\
\hline
2.20 & 
 $0.97\pm0.02$ & $1.05\pm0.05$ &
 $0.86\pm0.16$ & $0.75\pm0.10$ \\
1.99 & 
 $0.94\pm0.03$ & $1.07\pm0.04$ &
 $0.89\pm0.22$ & $0.68\pm0.12$ \\
\hline
\end{tabular}\end{center}
\caption{Thermal shifts in masses and mass splittings at $T=0.95T_c$;
$\mu_H$ denotes the screening length in the hadronic channel $H$, and
$m_H$ the mass at $T=0$. We find that $\mu_H=m_H$ within 95\% confidence
limits, indicating no significant thermal effects in the vector meson,
V and nucleon, $N_+$. Nor do we observe a thermal shift in the mass
splittings between the V and the AV (axial vector) mesons. However,
the splitting between the $N_+$ and its parity partner, $N_-$, changes
at finite temperature, and is a precursor to the restoration of chiral
symmetry before the QCD phase transition.}
\eet{rats}

Experiments with relativistic heavy-ion collisions are beginning
to probe the finer details of non-perturbative predictions of QCD
\cite{glmrx}. This is possible because fireballs produced in these
collisions come close to thermal equilibrium \cite{hrg}, and lattice QCD
techniques can be used to predict expectation values of observables under
these conditions \cite{tc}. The question of why the fireballs come into
equilibrium, \ie, why relaxation times are small \cite{heinz}, remains
outside the direct purview of lattice computations for now. Even so,
one can ask why thermodynamics is applicable \cite{koch}. The answer
can be found by testing whether any static correlation length in a QCD
medium, $\xi$, (which is the inverse of a screening mass, $\mu$)
is much smaller than the size of the fireball.

The only correlations which persist in the equilibrium thermodynamic
system are static spatial correlations. These correlation functions
are measured by introducing static probes into the equilibrium plasma
and measuring the response of the medium.  This response depends on
the quantum numbers carried by the probes; so one can classify static
correlators as glueball-like, meson-like and baryon-like probes with
the usual quantum numbers of these quantities corrected for the fact
that the static spatial symmetries are different from the Poincare
group. The group theory of these correlators was worked out for glueballs
\cite{gball0,rep0,gball1} and mesons \cite{rep1}. In \apxn{B} we extend
the representation theory to spin-1/2 baryons.

Meson-like screening masses have been studied in QCD in great detail
\cite{DK01,GRH,mtc,GGM01,Wis01,GGL01,refs,prasad}.  While we continue
these studies here, the main focus of this work is the baryon sector.
Baryon-like screening correlators were studied in the high-temperature
phase \cite{DK01,GRH}, but there have been no follow-up studies in recent
years, nor have these been extended to the low-temperature phase till now.
Since these provide important inputs to the study of baryon-number
fluctuations, which is of much interest in the experimental search for
the critical point of QCD, we reboot the study of these correlators.

Above $T_c$ we verify that baryonic screening lengths support the evidence
for weakly interacting quarks which has been gathered over the years
with meson screening correlators. Below, but very close to, $T_c$ we
find no evidence of any finite temperature effects on the masses of the
lowest mesonic resonances.  However, we find that in the nucleon sector
the mass of the opposite parity nucleon moves closer to the ground state
(see \tbn{rats}). These observations not only constrain models of quantum
hadrodynamics \cite{rappwambach}, but could also have implications for
the analysis of heavy-ion collision data.

The paper is organized as follows. In \scn{TD} we discuss technical
details related to the simulation, inversion and analysis, including the
details of interpolating operators used and the asymptotic fit forms. In
\scn{mesons} and \scn{baryons} we display the results in detail. \scn{Conc}
summarizes and concludes the work done in this paper. Various technical 
details are presented in appendices.

\section{Runs and measurements \label{s:TD}}

\bet[tbh]
\betb{|c|c|c|c|c|c|}\hline
$T/T_c$ & $N_\tau\times N_s^3$ & $\beta$ & $N_{conf}$ & $c_{\scriptscriptstyle{SW}}$ & $\kappa$ \\ \hline
0    & $32\times16^3$ & 6.03  & 71 & 1.7333 &  0.1345, 0.1347 \\
0.95 & $8\times32^3$  & 6.03  & 94 & 1.7333 &  0.1345, 0.1347 \\
1.5  & $8\times32^3$  & 6.332 & 67 & 1.5667 &  0.1345, 0.1350, 0.1355 \\
\hline\eetb
\caption{The simulation and measurement parameters used in this
work.  $\beta$ is the bare coupling for the Wilson gauge action,
$c_{\scriptscriptstyle{SW}}$ is the clover coefficient (determined
non-perturbatively in \cite{alpha}), and $\kappa$ is the hopping parameter
in the improved Wilson-Dirac operator.}
\eet{sim_det}

We measured correlation functions of operators with meson and baryon
quantum numbers built out of clover-improved Wilson quarks \cite{SW01}
in zero and finite temperature ensembles of configurations generated
using the Wilson gauge action. The finite temperature configurations were
generated on lattices with lattice spacing $a=1/(8T)$ with $T=0.95T_c$
and $1.5T_c$. The spatial size of the lattice, $L$, was tuned such
that $LT=4$. We use the notation $V=L^3$, $V_4=V/T$, $N_t=1/(aT)$ and
$N_s=L/a$. Zero temperature measurements were carried out on $(2L)\times
L^3$ lattices with coupling $\beta = 6.03$, which corresponds to
the same lattice spacing as the finite temperature run below $T_c$
\cite{SG01,BO01}. We will describe later that this meant that $L$ was
8.4 times the Compton wavelength of the lightest state on the lattice,
as a result of which finite volume effects are under control.  A zero
temperature study at $\beta = 6.332$, corresponding to the thermal
ensemble above $T_c$ was not  made, but the corresponding choice of
$\kappa$ was made so that we could use a previous study \cite{BLT01}.
Details of the simulation and measurement parameters are tabulated in
\tbn{sim_det}.  With the choices of $\kappa$ listed for $\beta=6.332$, the
values of $m_\pi/T_c$ are 1.52, 2.58, and 3.31 respectively. For the other
parameters, the results of the mass measurements are described later.

The gauge configurations were generated by a combination of one heat
bath and three  overrelaxation updates per Monte Carlo (MC) step. The
first  2000 MC steps were discarded to ensure equilibration of  the
lattices. Subsequently, measurements were performed once in 200 MC
steps, which is  approximately 30 to 100 times the autocorrelation
time of the action. As a result, it took 4 CPU hours to generate a
single configuration on our largest lattice. 

The measurement of hadronic correlation function requires the inversion
of the Dirac operator. This was done by a conjugate gradient algorithm
which stopped when the residual dropped below $(V_4/a^4) \times 10^{-16}$.
Typically this was found to take about $2500$ conjugate gradient steps,
and about 6 hours of CPU time, for the smallest pseudo scalar mass
used in our calculations below $T_c$. The computations above $T_c$
are substantially cheaper; the inversions  converged in approximately
0.67 CPU hour for our lightest pseudo scalar mass.

The zero momentum correlation functions are
\beq 
   S_\had(t) = \sum_\x \langle H^\dag(\x,t)H(\mathbf 0,0) \rangle, 
    \qquad{\rm where}\qquad
   H(\x,t) = \bar{\psi}(\x,t)\Gamma_\had\psi(\x,t),
\eeq{meson}
where $\psi(\x,t)$ is the quark field at time $t$ and spatial
point $\x$ (we will also use the component notation $\x_1=x$, $\x_2=y$
and $\x_3=z$).  The sum over all spatial sites of the point-to-point
correlator projects on to zero spatial momentum. All Dirac, flavour,
and colour indices are summed. $\Gamma_\had$ is an appropriate Dirac,
flavour matrix which gives the quantum numbers of the meson $H$. In actual
practice, we replaced the point source of \eqn{meson} by a wall source,
in order to control the statistics better. This required gauge fixing;
we fixed to the Coulomb gauge.  The effect on the extraction of screening
masses is discussed in \apxn{specfn}.

We used $\Gamma_\PS=\gamma_5$ for the isovector pseudoscalar (PS),
$\Gamma_\SC=1$ for the isovector scalar (S), $\Gamma_\V=\gamma_i$
for the vector (V) and $\Gamma_\AV=\gamma_i\gamma_5$ for the isovector
axial vector (AV) channels. For the $T=0$ measurements, we considered
propagation in the time direction as shown in eq.\ (\ref{meson}), and
since all the orthogonal directions are equivalent, we summed the V and
AV propagators over the three polarizations $i=1,2,3$.  For the screening
correlators all the polarizations of the V and AV are not equal, and we
summed over only $i=1$ and 2.

The lattice mesonic currents need to be renormalized to connect them 
with the continuum currents. For quarks of mass $m_q$, the lattice 
operators, $H(\x,t)$, are multiplicatively renormalized to
\beq
   Z_\had(g^2) \Big(1+b_\had(g^2) a m_q \Big) H(\x,t),
\eeq{eq:renorm}
where $a$ is the lattice spacing, $g^2 = 6/\beta$, and $am_q =
(\kappa - \kappa_c)/2$.  We use the results in the $\overline{MS}$
scheme where the renormalization constants are determined at the scale
of $1/a$.  For non-perturbatively improved clover fermions, the factors
$Z_{\V,\AV}$ and $b_\V$ have
been calculated non-perturbatively \cite{alphap},  where the following
interpolating formulae are found:
\beq
   Z_\V = \frac{1-0.7663g^2+0.0488g^4}{1-0.6369g^2}, \quad
   Z_\AV = \frac{1-0.8496g^2+0.0610g^4}{1-0.7332g^2}, \quad
   b_\V = \frac{1-0.6518g^2+0.1226g^4}{1-0.8467g^2}.
\eeq{eq:nprenorm}
For the other coefficients, we use the expression from the one-loop, tadpole
improved perturbation theory \cite{lepage}:
\beq
   Z_{\PS,\SC} = u_0\Big(1 - z_{\PS,\SC} \tilde g^2 \Big),
\eeq{eq:prenorm}
where $u_0$ is the tadpole factor, $\langle P \rangle = u_0^4$,
$\tilde{g}^2 = g^2/u_0^4$, and the one-loop coefficients $z_{\PS,\SC}$
as obtained from  \cite{gockeler} are $z_\PS = 0.107-0.019 \tilde c_\SW
+ 0.017 \tilde c_\SW^2$ and $z_\SC = 0.026 + 0.065 \tilde c_\SW - 0.012
\tilde c_\SW^2 $ where $\tilde c_\SW = u_0^3 c_\SW$.  The one-loop order
tadpole-improved forms of $b_\had$ can be written as
\beq
   b_\had = \frac1{u_0}\Big(1+b^1_\had\tilde g^2\Big).
\eeq{eq:massdeprenorm}
where one obtains from the calculations of \cite{sint} the values
$b^1_\PS=0.109$, $b^1_\SC=0.070$, and $b^1_\AV=0.069$, computed for the
tree-level $\tilde c_\SW=1$.

Masses and screening masses can be obtained without renormalization. They
were estimated, as usual, by fitting to an assumed cosine-hyperbolic
form, and looking for agreement with effective masses. These imply
assumptions about the spectral function at zero and finite temperatures,
and we discuss these in more detail in \apxn{specfn}. Statistical errors
on masses are obtained by a jackknife procedure. The propagation of
statistical errors is done by jackknife when they could be correlated,
and by adding in quadrature when they are independent.  The scalar
propagator has strong quenching artifacts. This prevents the use of
these correlators in estimating masses in the usual way (see \apxn{scl}).

Chiral symmetry restoration in the high temperature phase of QCD is signalled
by pairwise equality of correlators which are related to each other by parity.
The observation of some breaking of this symmetry has been an issue of some
interest recently \cite{refs}. Here we introduce a new measure for this
symmetry---
\beq 
   R_\had = \frac1{N_t-1}\sum_{t=1}^{N_t-1}
    \frac{\langle S_{\had_+}(t)-S_{\had_-}(t)\rangle}{
          \langle S_{\had_+}(t)+S_{\had_-}(t)\rangle}\,,
\eeq{chisym_meas}
where $S_{\had_\pm}(t)$ are parity partner correlators (for the screening
correlators $t$ is replaced by $z$ and $N_t$ by $N_s$), and the angular
brackets are averages over the gauge ensemble. The correlators at $t=0$
are left out of the sum to avoid problems with time doublers. We will
use the convention of taking the positive parity partner for the label
H. When chiral symmetry is broken we expect $R_\had\simeq{\cal O}(1)$.

For the nucleon operator we used
\beq 
   N_\alpha(\x,t) = \varepsilon_{abc} (C\gamma_5)_{\beta\delta} 
      \psi^a_\alpha(\x,t) \psi^c_\beta(\x,t) \psi^b_\delta(\x,t),
\eeq{nucleon}
where $\alpha$, $\beta$, and $\gamma$ are Dirac indices, $a$, $b$,
and $c$ are colour indices, $\varepsilon$ is the Levi-Civita symbol,
and $C$ is the charge conjugation operator. The projection of the
correlation function on to vanishing spatial momentum is performed by
the usual means of summing over $\x$ at $T=0$ where we imposed periodic
boundary conditions in all directions. However, at finite temperature,
anti-periodic boundary conditions must be imposed on quark fields in
the Euclidean time direction. As a result, the screening correlator must
be projected to the lowest Matsubara frequency \cite{DK01}. The parity
projection operators for the correlator whose propagation is measured
in the direction $\mu$ is 
\beq
   \mathcal{P}^{\mu}_{\pm} = \frac12(1 \pm \gamma^{\mu}).
\eeq{nucl_parity_proj} 
The group theory relevant for screening correlators is given in \apxn{B}.

The zero momentum correlators of the two parities of the nucleon can be
fitted to the behaviour expected of a Wilson fermion---
\beqa 
\nonumber
   S_\Np(t) &=& c_+\exp{[-\bar\mu_\Np t]}
            +c_-\exp{[-\bar\mu_\Nm(N_t- t)]}\\
   S_\Nm( t) &=& c_-\exp{[-\bar\mu_\Nm t]}
            +c_+\exp{[-\bar\mu_\Np(N_t-t)]}.
\eeqa{Wilfer_cor}
Because of the admixture from opposite parity states, these correlators change
sign across the middle of the lattice. In this paper we used the absolute value
of the correlators above. We note that $R_\N$ does not require knowledge of
$Z_\N$ because the parity partners are generated by the same lattice operator.
For the $T=0$ measurement, where 
$\bar\mu_{\scriptscriptstyle N^\pm}=m_{\scriptscriptstyle N^\pm}$. For the
finite temperature measurement where we replace $t$ by $z$, we have 
\beq 
   \mu_{\scriptscriptstyle N^\pm}^2 =
   \bar\mu_{\scriptscriptstyle N^\pm}^2 - \sin^2(\pi /N_t),
\eeq{mat_correction}
when we use the prescription of \cite{DK01}.

\section{The meson sector \label{s:mesons}}

\bet[tbh]
\betb{|c|ccc|}\hline
$\kappa$ & $m_\pi/T_c$   & $m_\V/T_c$      & $m_\AV/T_c$\\
\hline
$0.1345$ & $2.20\pm0.03$ 
         & $3.55\pm0.06$ 
         & $5.6\pm0.4$ \\
$0.1347$ & $1.99\pm0.03$ 
         & $3.53\pm0.08$ 
         & $5.5\pm0.5$ \\
\hline\eetb
\caption{Meson masses in units of $T_c$ at $T=0$ and $\beta=6.03$.}
\eet{t0mes}

The analysis of meson masses at $T=0$ is absolutely standard. The results
are collected in \tbn{t0mes}. We choose to express all results in units
of $T_c$ for two reasons. First, because quenched QCD is not open to
experimental tests and hence quoting numbers in MeV units is based
on assumptions which cannot be tested. We prefer to quote ratios of
quantities which are computable in practice. Second, because in quenched
QCD the critical coupling is known with high precision, and $N_t=8$ is
within the scaling region \cite{scaling}, the statistical and systematic
errors involved in using this as a scale are completely under control.

\bet[tbh]
\betb{|c|ccc|}\hline
$m_\pi/T_c$ & H & $\mu_\had/T_c$ & $\mu_\had/m_\had$ \\
\hline
2.20 & PS & $2.18\pm0.02$ 
          & $0.99\pm0.02$\\ 
     & V  & $3.45\pm0.06$ 
          & $0.97\pm0.02$\\ 
     & AV & $5.18\pm0.09$ 
          & $0.93\pm0.06$\\ 
\hline
1.99 & PS & $1.97\pm0.02$ 
          & $0.99\pm0.02$\\ 
     & V  & $3.31\pm0.07$ 
          & $0.94\pm0.03$\\ 
     & AV & $5.1\pm0.1$ 
          & $0.92\pm0.08$\\ 
\hline\eetb
\caption{Meson screening masses, $\mu_\had$ at $T=0.95T_c$ in units of $T_c$
and the corresponding $T=0$ meson mass, $m_\had$.}
\eet{warmmes}

The analysis of most mesonic correlators below $T_c$ is equally
straightforward. The only subtlety has been mentioned earlier: since the
zero momentum screening correlator is measured for separations along
the $z$-direction, the three polarizations states of the V and AV are
two spatial and one temporal.  The temporal polarizations have behaviour
distinct from the spatial polarizations \cite{rep1}.  We measure the
screening masses of the spatial polarizations only.  Our main results
for the screening masses below $T_c$ are summarized in \tbn{warmmes}.
We discuss the relation between the screening mass and the pole mass
in \apxn{specfn}. Our results show that the pole mass of the mesons is
hardly affected by temperature.

\bet[tbh]
\betb{|c|cc|cc|}\hline
 &\multicolumn{2}{c|}{$T=0$}&\multicolumn{2}{c|}{$T=0.95T_c$}\\ \cline{2-5}
$m_\pi/T_c$ & $\eint/m_\pi$ & $\chi^2/dof$
            & $\eint/m_\pi$ & $\chi^2/dof$ \\
\hline
2.20 & $0.6\pm0.2$  & 1.92 
     & $-0.1\pm0.8$ & 0.25\\ 
1.99 & $0.5\pm0.2$  & 1.28 
     & $0.1\pm0.7$  & 0.57\\ 
\hline\eetb
\caption{$\eint$ in units of $m_\pi$ obtained by fitting the scalar
 correlator with the fit form in \eqn{QChPT_long_t_beh_sc}.}
\eet{fitsc}

In \apxn{scl} we discussed the reasons why scalar decays are visible
in quenched QCD, and how the
final-state interaction energy, $\eint$,
can be extracted.  The analysis involves fitting the functional form in
\eqn{QChPT_long_t_beh_sc} at both $T=0$ and for $T<T_c$.  The extracted
physical parameters are given in \tbn{fitsc}.
$\eint$ at $T=0$ is non-zero at the 95\%
significance level for both values of $m_\pi$ which we used.  While the
finite temperature measurements are statistically indistinguishable
from these, they are consistent with zero within errors, due to the
larger fit errors at $T>0$.  It is interesting to note that the central
values of $\eint$ at $T>0$ lie outside the $2\sigma$ errors of the $T=0$
measurement.

\bef[b]
\centering
\subfigure[]{\incfig{0.65}{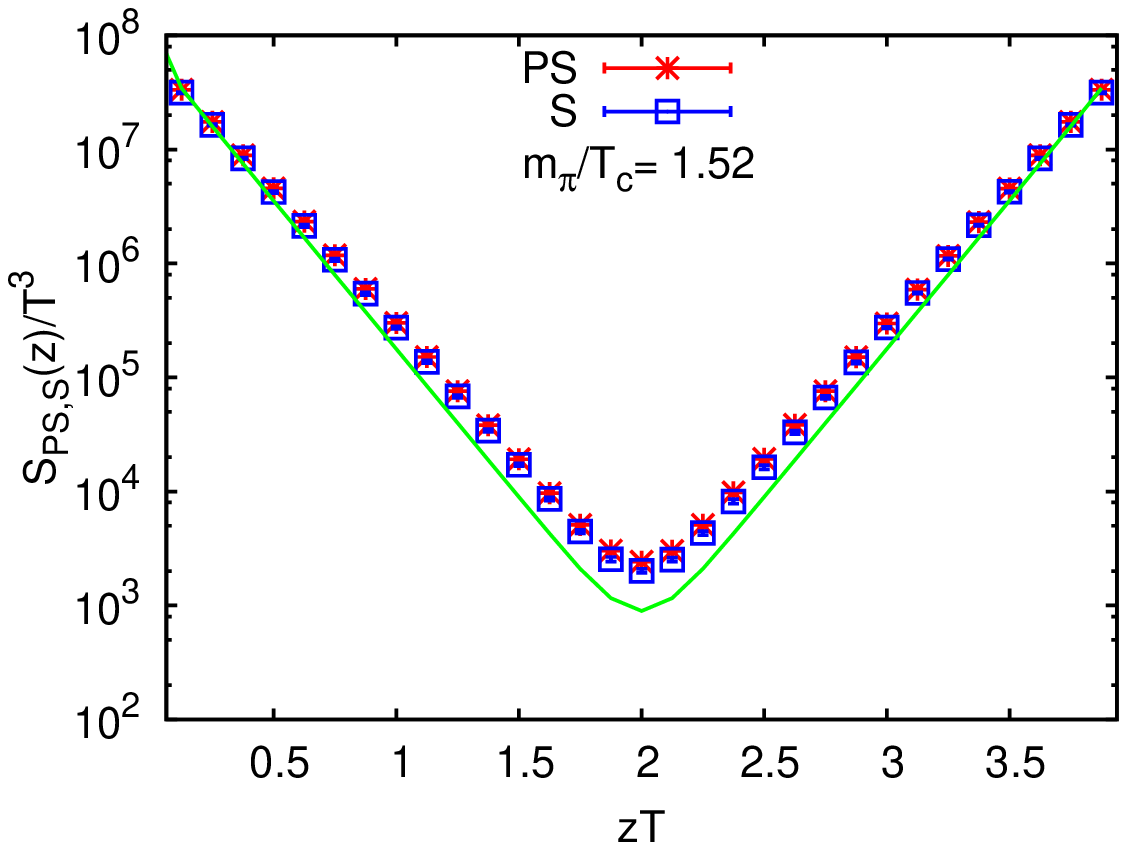}} \qquad
\subfigure[]{\incfig{0.65}{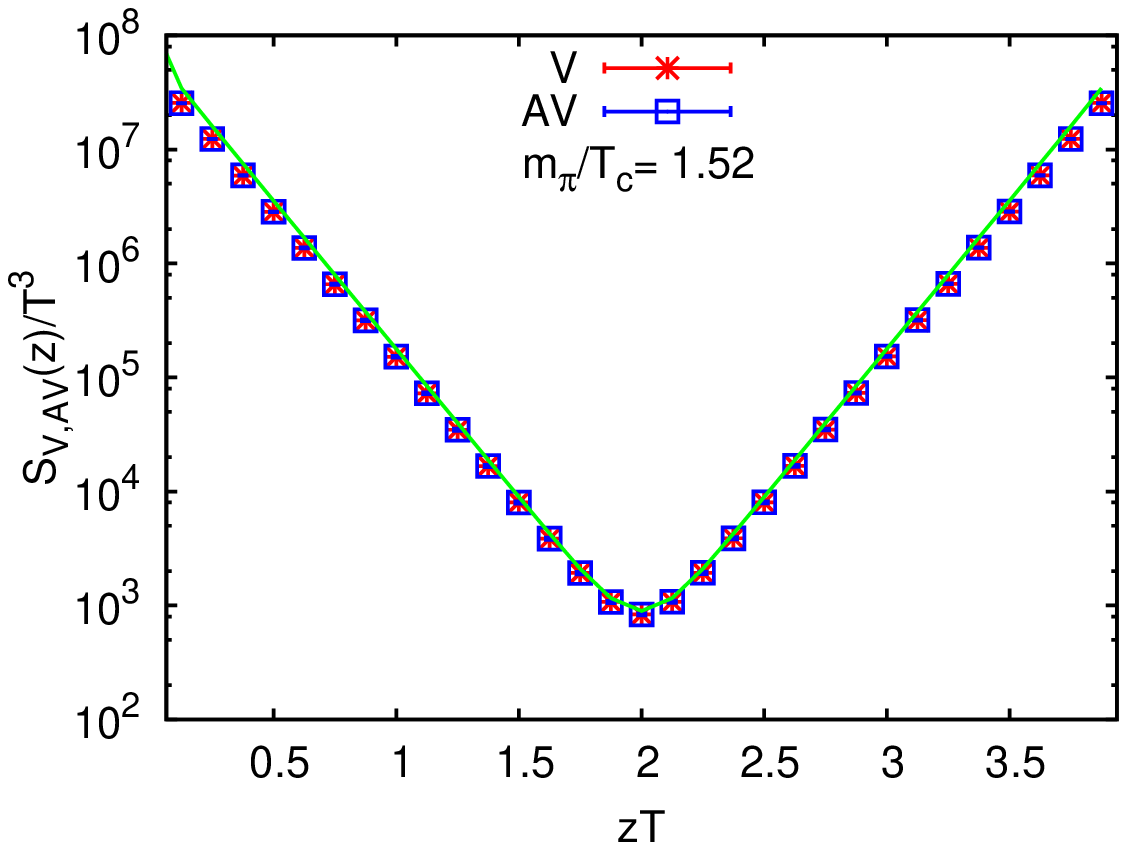}}
\caption{A signal of restored chiral symmetry above $T_c$ is that (a) the
PS and S correlators and (b) the V and AV correlators become degenerate.
The continuous curves are correlation functions in a FFT; in order to
remove trivial artifacts, these have been computed on a lattice of the
same size.}
\eef{chirest}

\bef
\incfig{0.65}{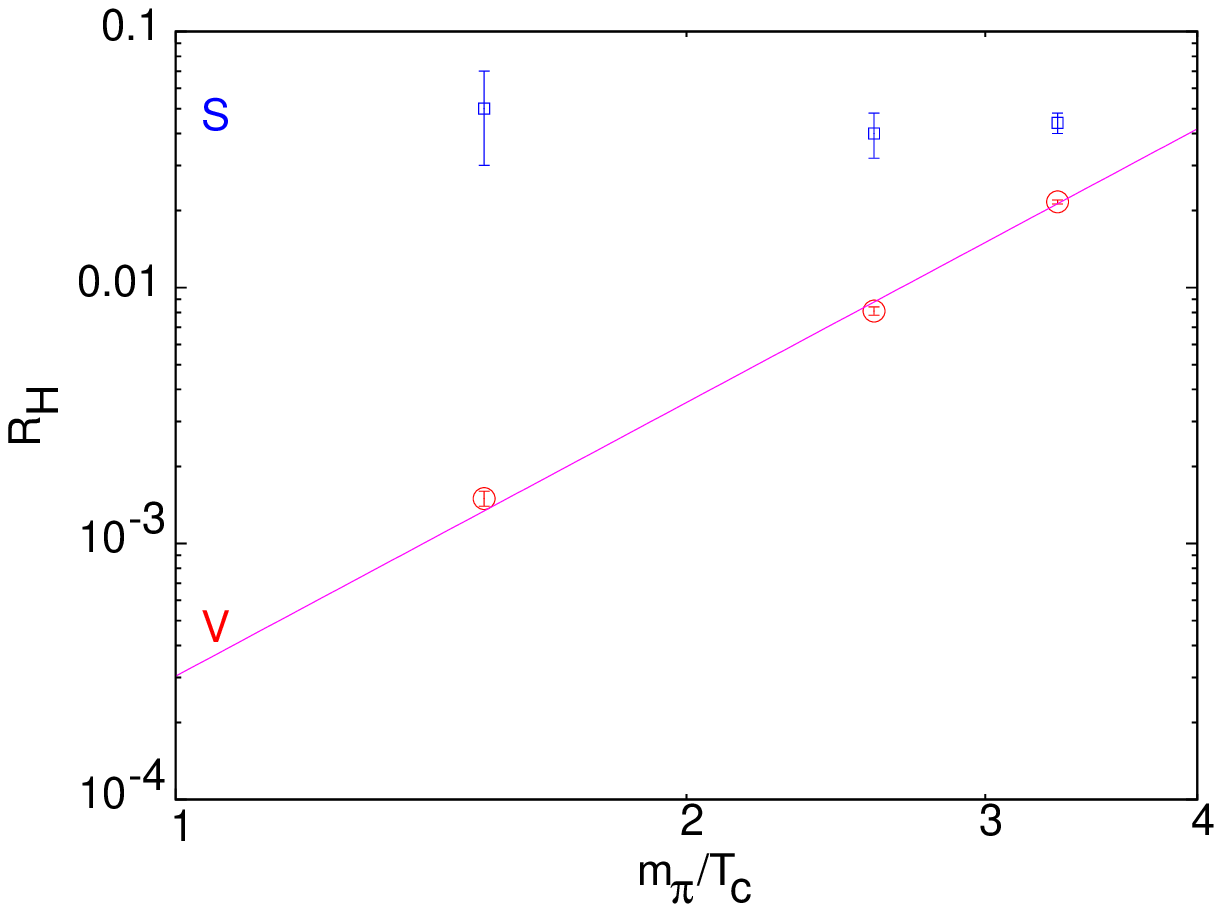}
\caption{Above $T_c$ $R_\SC$ is found to be almost independent of
$m_\pi/T_c$. However, $R_\V$ goes to zero as a power of $m_\pi/T_c$,
showing that correlation functions of the parity partners $V$ and $AV$
do become exactly degenerate in the limit.}
\eef{rhad}

The ${\rm SU}_L(2)\times{\rm SU}_R(2)$ chiral symmetry of the QCD vacuum
is restored above $T_c$. The anomalous ${\rm U}_A(1)$ remains broken
asymptotically, although its effects are seen to be small at temperatures
of $1.5T_c$. The most straightforward consequence of this is that the S/PS
and the V/AV correlators should become pairwise degenerate in the high
temperature phase. That this is the case is shown in \fgn{chirest}. The
figure also shows that the correlation functions are not too far from
those expected in a theory of non-interacting quarks (also called the
free field theory, FFT).

At $T=0$ in the V/AV sector we found $R_\V=0.73\pm0.02$; at
$T=0.95T_c$ this drops marginally to $R_\V=0.59\pm0.02$ for both the bare
quark masses we have used. This indicates that chiral symmetry remains
strongly broken up to $0.95T_c$.  A simple model of V/AV mixing below
$T_c$ was presented in \cite{rappwambach} using a mixing parameter
$\epsilon$, which can be adapted to our use by writing schematically
\beq
  S_\V(T)=(1-\epsilon)S_\V(0)+\epsilon S_\AV(0), \qquad{\rm and}\qquad
  S_\AV(T)=(1-\epsilon)S_\AV(0)+\epsilon S_\V(0).
\eeq{rw}
This gives $R_\V(T) = (1-2\epsilon)R_\V(0)$. Using the values quoted
above, we find $\epsilon=0.10\pm0.02$. In the unquenched theory
one may expect much larger values of this parameter below $T_c$
\cite{rappwambach}.  We observed that at $T=1.5T_c$ the value of $R_\V$
drops as a power of the quark mass, vanishing as $m_\pi$ vanishes (see
\fgn{rhad}). However, there is no dependence of $R_\SC$ on the quark mass.

We have extracted screening masses, $\mu_\had$ from the correlation
functions. We found that effective masses showed a good plateau which
agreed with fits, unlike previous experience with Wilson quarks at $T>T_c$
\cite{Wis01}. The reasons for this are explored in \apxn{specfn} where
some effective mass plateaus are also displayed.  Our results for the
screening masses are collected in \tbn{hotscrms}. The pairwise degeneracy
of the masses is quite evident. It is also evident that the screening
masses are close to that expected in a theory of free quarks. The
screening masses in the S/PS channel are 7--9\% smaller than that in
the free theory, whereas in the V/AV channel they are within 2--3\%
of the free theory.

\section{The baryon sector \label{s:baryons}}

\bef[b]
\subfigure[]{\incfig{0.65}{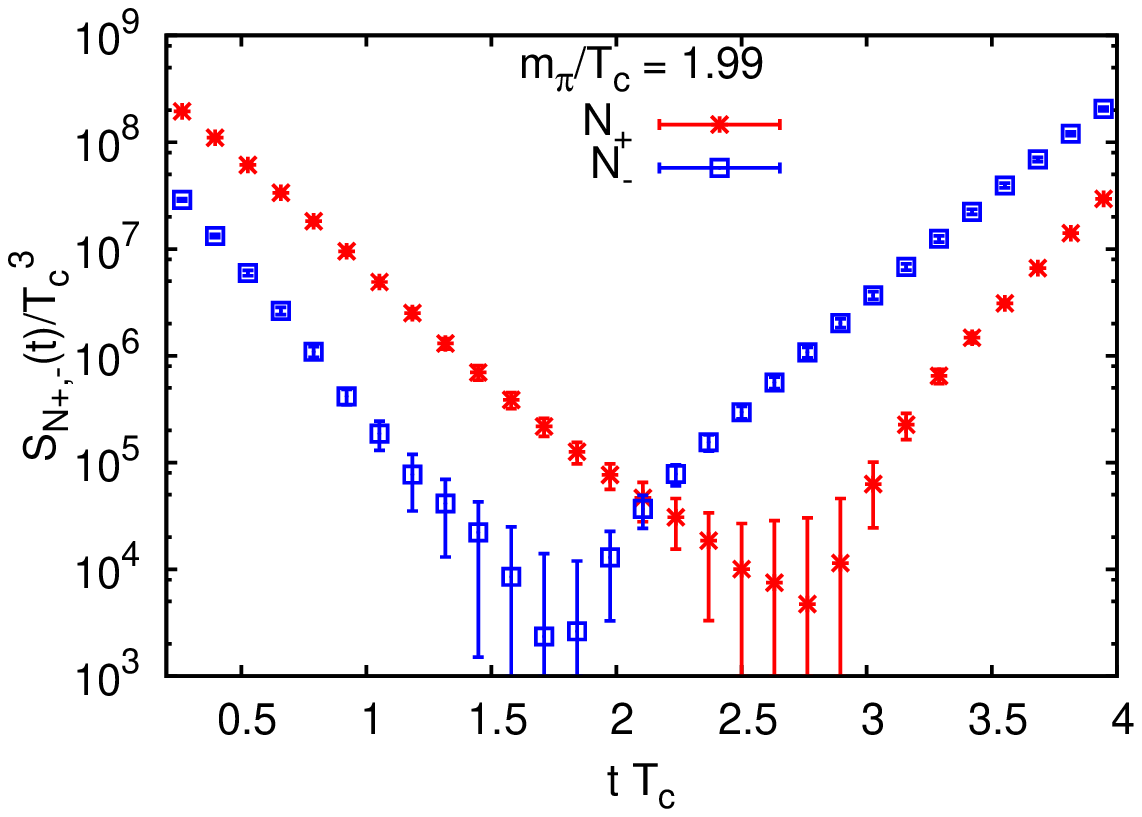}} \qquad
\subfigure[]{\incfig{0.65}{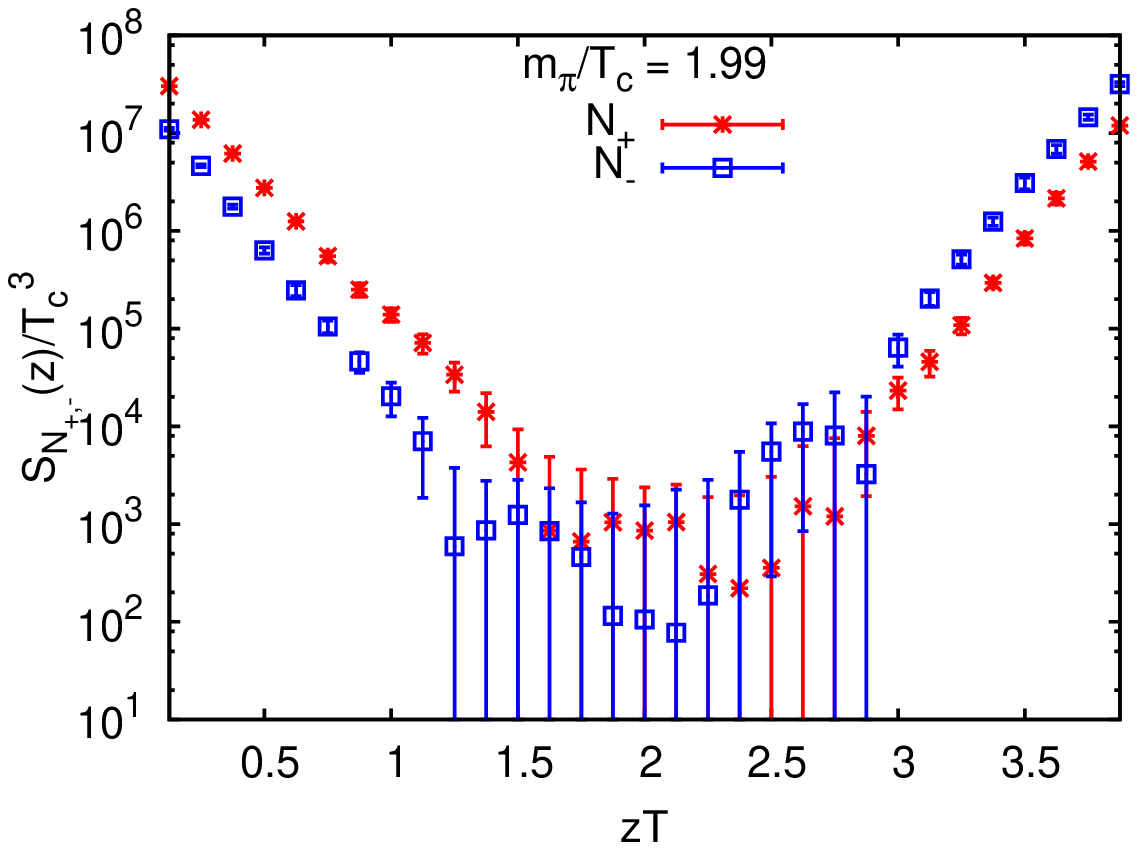}}
\caption{Nucleon correlators at (a) $T=0$ and (b) $T=0.95T_c$. Both sets are
asymmetric, showing that the nucleon, $N_+$ and its parity partner, $N_-$,
are not degenerate.}
\eef{ncor}

Analysis of baryon correlators requires more care than meson
correlators. This is because states of opposite parity contribute to
each parity projected correlator, as shown in \fgn{ncor}. As a result,
both the fitting procedure and the extraction of effective masses is more
complex than the analysis for mesons. In addition, the projection of the
screening correlator on to zero momentum requires a twist to compensate
for the thermal boundary condition.  The subsequent extraction of a
screening mass from the nucleon correlator also requires the subtraction
of \eqn{mat_correction}, resulting in additional loss of precision.

In \fgn{ncor} it would appear that the finite temperature correlators
are more nearly symmetric than those at $T=0$. Whether or not there
is an early onset of chiral symmetry restoration can be probed by
constructing the measure $R_\N$, defined by \eqn{chisym_meas}. At
$T=0$ we find $R_\N=0.88$ when $m_\pi/T_c=2.20$ and 0.89 at the lower
quark mass. At finite temperature these change to $R_\N=0.8$ and 0.83
respectively. Contrary to the visual impression created by \fgn{ncor},
once the covariances between the $N_\pm$ correlators are accounted for,
the correlation functions themselves do not show any tendency towards
early restoration of chiral symmetry. In fact if we adapt the model
of \eqn{rw} to this case, we find a mixing parameter $\epsilon=0.04$
in the nucleon sector, which is even smaller than that found for the
V/AV at the same temperature.

\bet[tbh]
\betb{|c|c|c|c|c|}\hline
$m_\pi/T_c$ & $m_\Np/T_c$ & $\mu_\Np/T_c$ & $m_\Nm/T_c$ & $\mu_\Nm/T_c$ \\ \hline
2.20 & $5.10\pm0.10$ 
     & $5.3\pm0.2$ 
     & $6.8\pm0.4$ 
     & $6.6\pm0.4$ \\
1.99 & $4.9\pm0.1$  
     & $5.28\pm0.14$ 
     & $6.7\pm0.5$ 
     & $6.4\pm0.7$ \\ 
\hline\eetb
\caption{Nucleon masses at $T=0$ and screening masses at $T=0.95T_c$.}
\eet{nuclfit}

\bef
\subfigure[]{\incfig{0.65}{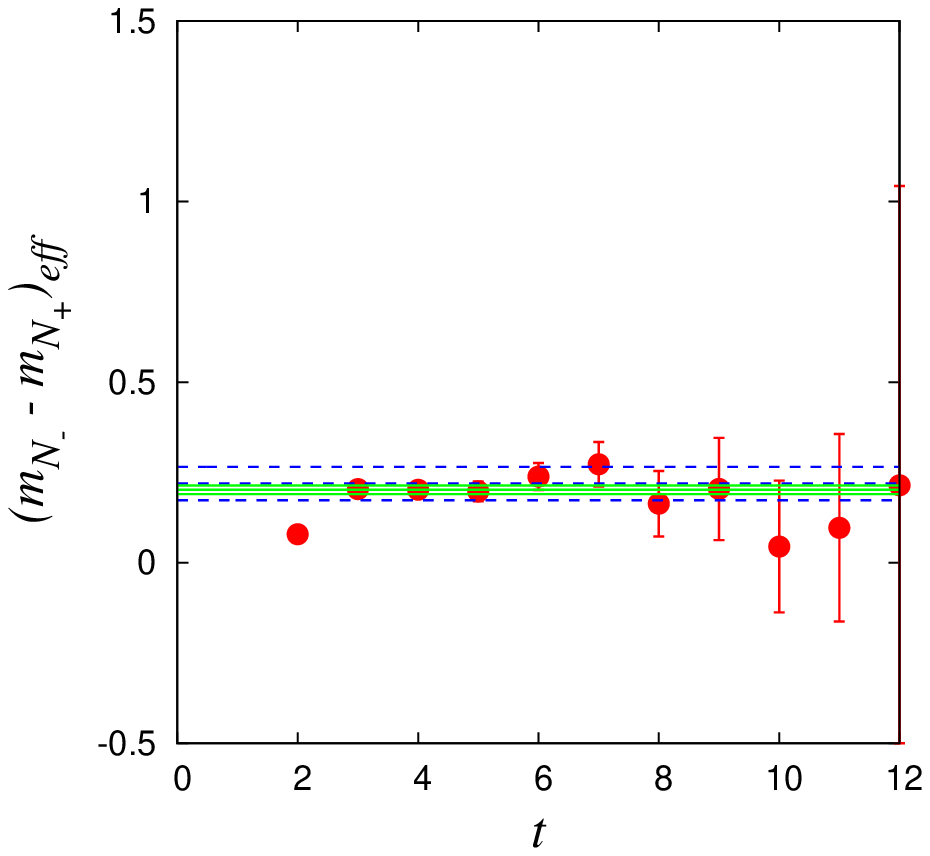}} \qquad
\subfigure[]{\incfig{0.65}{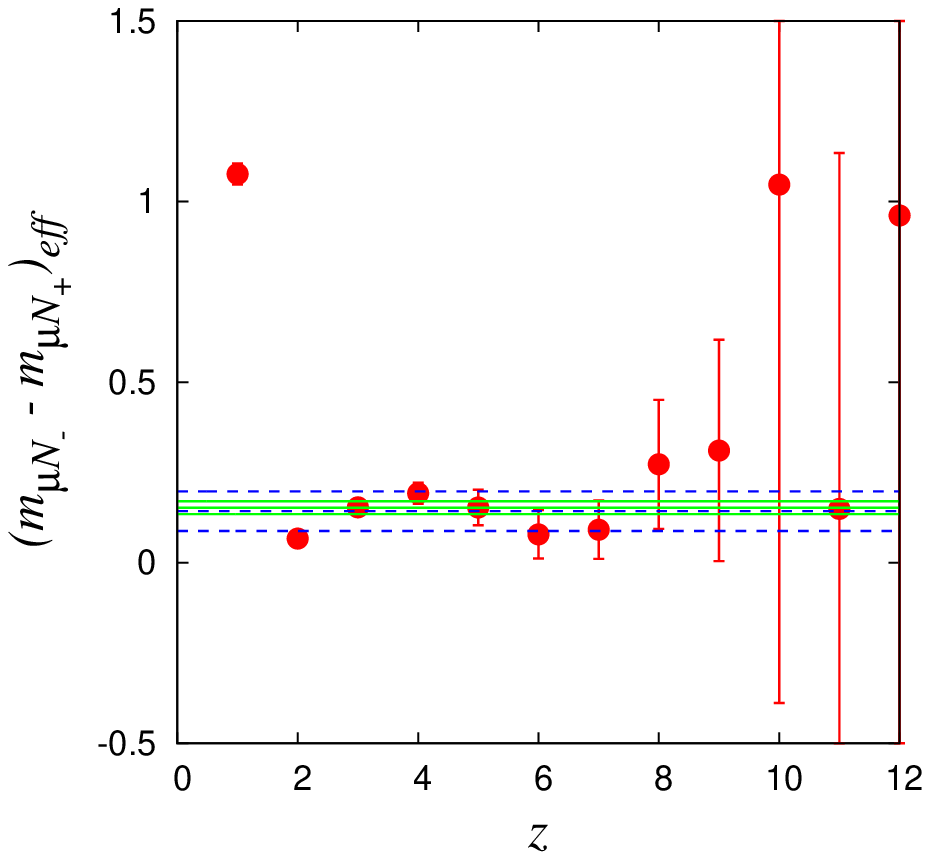}} \\
\subfigure[]{\incfig{0.65}{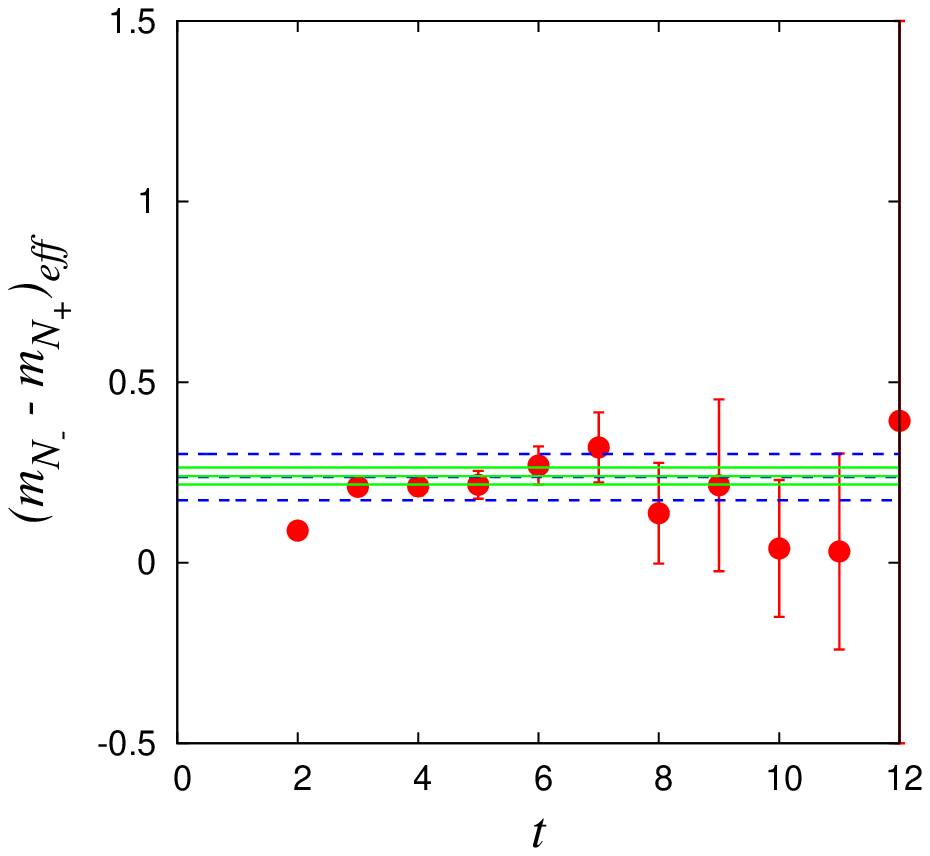}} \qquad
\subfigure[]{\incfig{0.65}{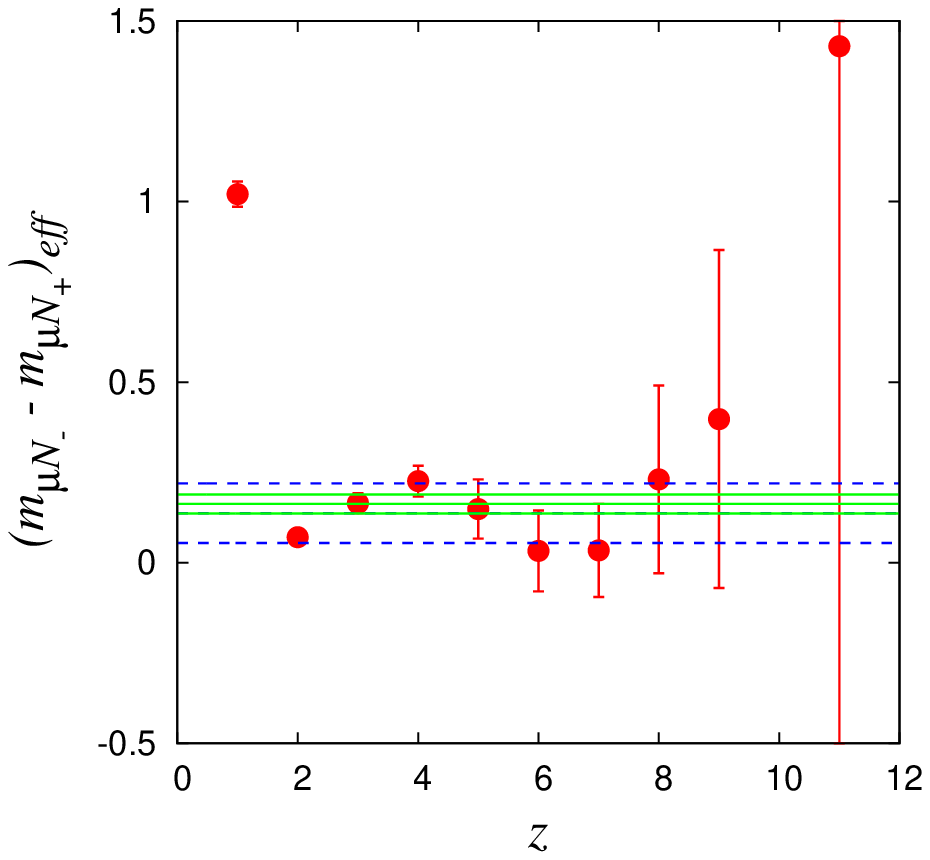}}
\caption{The analysis of mass splittings between $N_+$ and $N_-$ for
$m_\pi/T=2.20$ at (a) $T=0$ and (b) $T=0.95T_c$ and for $m_\pi/T=1.99$
at (c) $T=0$ and (d) $T=0.95T_c$. The data points are effective masses,
the band between the dashed lines is obtained by taking the difference of
screening masses. The band between the full lines is obtained by
direct fits to the splitting between the two masses, and gives the results
shown in \eqn{split}.}
\eef{split}

The results of fitting the masses is shown in \tbn{nuclfit}. Here we indeed
find some precursor effects of chiral symmetry restoration in the form of a
thermal shift in the mass splitting between the baryon and its parity
partner. We measured the splitting by the ratio of the correlators,
$S_\Np/S_\Nm$, and extract the mass difference, $\Delta m$, by fitting
to an exponential form $\exp(-\Delta mt)$. Since this method takes care
of covariances between the correlators $S_\Np$ and $S_\Nm$, we expect
to control statistical errors better. The fit can be cross checked
against the equivalent of effective masses for the ratio. Such checks
are exhibited in \fgn{split}. The resulting values are
\beq
   \frac{m_\Nm-m_\Np}{T_c}
   = \begin{cases} 1.53\pm0.09& (m_\pi/T_c=2.20)\\
                   1.82\pm0.17 & (m_\pi/T_c=1.99) \end{cases}
   \quad{\rm and}\quad
   \frac{\mu_\Nm-\mu_\Np}{T_c}
   = \begin{cases} 1.15\pm0.14 & (m_\pi/T_c=2.20)\\
                   1.24\pm0.20 & (m_\pi/T_c=1.99) \end{cases}.
\eeq{split}
The thermal effect is significant, and leads to the results quoted in
\tbn{rats}. This is the most definite evidence to date about precursor
effects to chiral symmetry restoration at $T_c$.

Above $T_c$ the correlators for $N_\pm$ become symmetric and degenerate.
We find $R_\N=0.230\pm0.005$ when $m_\pi/T_c=3.31$, $R_\N=0.146\pm0.006$
for $m_\pi/T_c=2.58$ and $R_\N=0.076\pm0.009$ for $m_\pi/T_c=1.52$. These
lead to a vanishing of $R_\N$ as a power of $m_\pi/T_c$.  Chiral symmetry
restoration is also seen in the fitted screening masses, displayed in
\tbn{hotscrms}. As in the meson sector, the screening masses are within
3--4\% of those expected in a theory of free quarks.

\section{Summary and conclusions \label{s:Conc}}

\bet[b]
\betb{|c|cccccc|}\hline
$m_\pi/T_c$ & $\mu_\PS/T$ & $\mu_\SC/T$ & $\mu_\V/T$ & $\mu_\AV/T$
            & $\mu_\Np/T$ & $\mu_\Nm/T$ \\
\hline
1.52 & $5.54\pm0.02$ & $5.59\pm0.04$ 
     & $5.88\pm0.02$ & $5.89\pm0.02$ 
     & $8.72\pm0.10$ & $8.68\pm0.10$ \\
2.58 & $5.56\pm0.02$ & $5.59\pm0.02$ 
     & $5.90\pm0.02$ & $5.91\pm0.02$ 
     & $8.75\pm0.10$ & $8.73\pm0.10$ \\
3.31 & $5.61\pm0.02$ & $5.65\pm0.02$ 
     & $5.95\pm0.02$ & $5.96\pm0.02$ 
     & $8.82\pm0.08$ & $8.77\pm0.09$ \\
\hline
\eetb
\caption{Hadron screening masses at $T=1.5T_c$ for three different masses
of quarks. In FFT all the mesonic (baryonic) screening masses are 
expected to be 5.99 (8.44), 5.995 (8.456) and 6.01 (8.48) on our lattices,
when the bare quark mass is tuned to give $m_\pi/T_c=1.52$, $2.58$ and $3.31$ 
respectively.}
\eet{hotscrms}

\bef[bht]
\incfig{0.8}{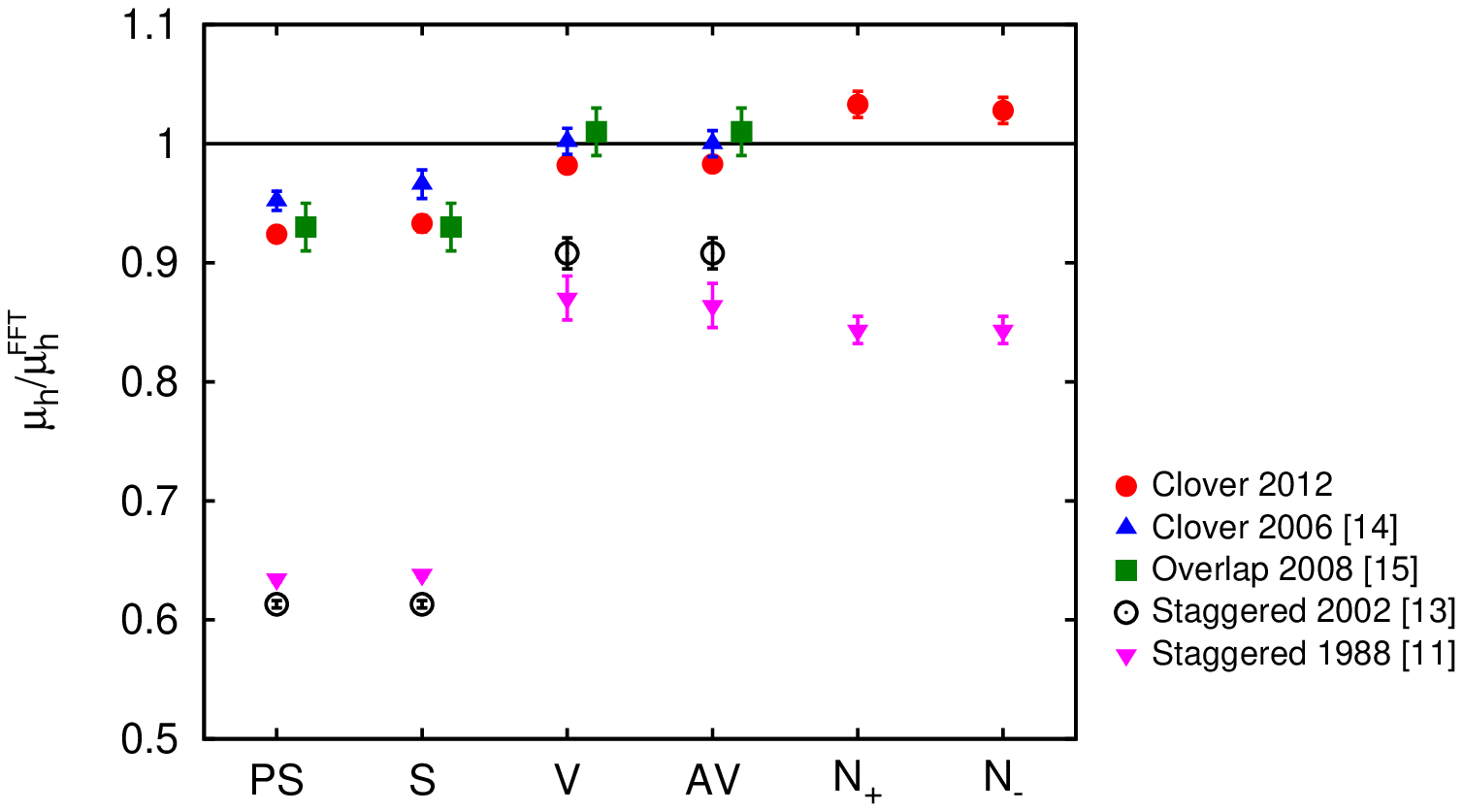}
\caption{The ratio of screening masses measured at $T=1.5T_c$ in quenched
QCD with those in FFT. Although all the results are not taken for exactly the
same quark mass, the effects of the quark mass are very small (less
than 1\% in this study). The results of these present computations are
in agreement, within 95\% confidence limits, with those from earlier
computations with clover \cite{Wis01} and overlap \cite{GGL01} quarks.
However, the results from earlier computations with staggered quarks
\cite{GRH,GGM01} differ considerably, being further away from the
limit of free fermions.}
\eef{freetheory}

We have studied thermal correlations in quenched QCD in channels
with various hadronic quantum numbers. The correlation functions are
consistent with the usual picture of the QCD phase diagram--- below $T_c$
long-distance correlations are mediated by hadrons and above $T_c$ these
are mediated by weakly interacting quarks. It is approximately true that
there are no thermal effects below $T_c$ (see \tbn{rats}) and that above
$T_c$ the effects seen are consistent with almost non-interacting quarks
(see \tbn{hotscrms}). However, there are several details which may impact
our understanding of the thermal behaviour of QCD.

Above $T_c$, at the temperature $T=1.5T_c$, we found strong signals
of approximate chiral symmetry restoration in the near degeneracy of
screening masses of hadronic parity partners, evident in \tbn{hotscrms}.
In addition we found, as before \cite{mtc,GGM01,Wis01,GGL01,refs}, that
the correlation functions and screening masses are not far from those
obtained in FFT, which is a model of non-interacting quarks.

Since there is persistent confusion in the literature regarding the
S/PS channels, we made a survey of the literature and found systematic
effects which we summarize in \fgn{freetheory}.  This work uses the
smallest lattice spacing used in a quenched study of screening to date,
and agrees with previous work using clover quarks \cite{Wis01}. It
also agrees with results of studies using overlap quarks at smaller
lattice spacing \cite{GGL01}. However, it seems that studies with na\"ive
staggered quarks in the quenched approximation always indicate stronger
deviations from FFT than any of these above studies.  Since studies
of screening with dynamical staggered quarks also claim such effects
\cite{mtc,refs}, it is worth examining cutoff effects with staggered
quarks more accurately in this context.

Even if the screening masses are close to that in the free theory at
$T=1.5T_c$, there are indications that there remain interesting physics
in correlations. The deviation from FFT is the opposite of that seen in
weak-coupling theory \cite{laine}.  Using the measure $R_\SC$, $R_\V$ and
$R_\N$ of \eqn{chisym_meas}, we examined correlations in parity partner
channels as a function of the quark mass (parametrized through the $T=0$
value of $m_\pi$). We found that as the quark mass is decreased $R_\V$
and $R_\N$ seem to vanish, as expected above $T_c$. However, $R_\SC$
does not depend on the quark mass, and remains non-zero. This still
does not resolve the question of the relative importance of the multiple
possible origins of this effect \cite{refs}, that will depend on future
temperature scans.

As highlighted in \tbn{rats}, mesons and the nucleon ground state are
indeed unaffected by thermal effects below $T_c$. We also find that the
splitting of the vector and axial vector meson masses is unaffected by
raising the temperature to $T=0.95T_c$. However, the opposite parity
excitation of the nucleon is seen to move closer to the ground state. A
closing of the mass gap between these two would signal chiral symmetry
restoration, and that is expected to occur only at $T_c$.  This is the
first observation of precursors of the phase transition in the closing of
the mass gaps of parity related states below $T_c$.  No such effects have
been seen earlier either in the glue sector or with quarks, nor do we see
any effect in the V/AV channels here.  Since the parameters $\epsilon$
introduced in \eqn{rw} is much smaller in the quenched theory than
expected in QCD with dynamical quarks, perhaps one can expect the change
in the mass gap to be more significant in the full theory. 

A decreased splitting between the nucleon and its resonances implies that
in a hadronic medium the interactions mediated through pions may keep the
resonances in chemical equilibrium even after other baryons freeze out.
This could result not only in changes in the net yield of nucleons,
as seen at the LHC, but also possibly in the momentum distribution
of nucleons.  These implications clearly call for follow up lattice
studies of nucleons below $T_c$.

{\bf Acknowledgements}: The computations reported here were performed on
the computing facilities of the ILGTI and the Department of Theoretical
Physics, TIFR. We would like to thank Ajay Salve and Kapil Ghadiali
for technical support. PM would like to acknowledge the Council of
Scientific and Industrial Research (CSIR) for financial support through
the Shyama Prasad Mukherjee fellowship. JM would like to acknowledge the
hospitality of the Department of Theoretical Physics, TIFR, where he was
a visiting fellow when he started working on this paper.

\appendix

\section{Splitting of Wilson-Dirac fermions at finite temperature\label{s:B}}

The point group of a \textit{z}-slice of the lattice at finite
temperature is the group $D_4$, which has $8$ elements in $5$
conjugacy classes. These classes are represented by the operators:
the identity, $\pi/2$ rotations about the \textit{t}-axis, $R_{xy}$,
$\pi$ rotations about the \textit{t}-axis, $R_{xy}^2$,  $\pi$ rotations
about the \textit{y}-axis, $R_{xt}^2$, and $\pi$ rotations about the
\textit{x+y}-axis, $R_{x+y}^2 = R_{xy}R_{xt}^2$. Since the finite
temperature theory is realized by periodic boundary conditions in the
Euclidean time direction for bosons, the symmetry group for pure gauge
operators is $D_4^h = D_4 \times Z_2(T)$ where $T$ is the reflection
operator in the $xy$ plane \cite{gball1}. Since fermions are realized
at finite temperature by anti-periodic boundary conditions, the $Z_2$
factor is not part of the symmetry group \cite{rep1}.

Acting on Dirac spinors, one can see immediately that
\beq
R_{xy}=e^{i\pi\sigma_{12}/4} = \frac1{\sqrt{2}}(1+i\sigma_{12}), \qquad  R^2_{xy} = i\sigma_{12}, \qquad R^2_{xt} = i\sigma_{14},
\eeq{dirac}
where $\sigma_{\mu\nu} = [\gamma_{\mu},\gamma_{\nu}]/2$ and $\gamma_{\mu}$ are 
the Euclidean Dirac gamma matrices. In the Euclidean chiral representation of 
the Dirac matrices, these three operators are block diagonal. The identity is 
always block diagonal, and $R^2_{x+y}$ is also block diagonal since both 
$R_{xy}$ and $R^2_{xt}$ are. As a result, under the action of $D_4$, the 
$4$-component Dirac spinor decomposes into two independent two-component 
objects, each corresponding to a specific helicity.

The $T=0$ parity operator is $P=(1+\gamma_4)/2$. Since this is (block) 
off-diagonal in the chiral basis, it mixes different helicity components. In 
the chiral symmetry restored phase, the two helicity components give 
degenerate correlation functions and one of the parity eigenstates
vanishes.

\section{Screening Correlator and the spectral function \label{s:specfn}}

Here we compile some relevant formulae for understanding the screening 
correlation function. We take the $z$ direction as the direction
of propagation, and examine correlators $S(z)$ with $z$ large.

The screening correlator is easy to understand in terms of the transfer
matrix of a $z$ slice, and is therefore perfect for understanding the 
symmetries of the finite temperature transfer matrix. Information
about thermal change of properties of hadrons, on the other hand, is
easier to obtain from the real-time retarded correlator \cite{lebellac}
\beq
 G^R(x;T) = i \theta(x^0) \langle \left[ H(x), H(0) \right] \rangle_T ,
\eeq{retarded}
where $\langle \rangle_T$ denotes thermal averaging at temperature $T$. 
The imaginary part of $G^R(p, T)$, the Fourier-transformed retarded correlator, 
is the spectral function $\sigma(p, T)$. A stable mesonic state, like the pion,
contributes a term 
\beq
 \sigma_\had(p) \simeq A_\had \epsilon(p_0) \delta(p^2-m_\had^2). 
\eeq{stable}
Here $A_\had$ depends on the details of the operator. In particular, 
in our case of mesonic operators, $A_\had$ depends on the nature of the source, 
and will be different for the wall and point sources. On the other hand, 
$m_\had$ is the property of the hadronic state.

At temperatures $T$ sufficiently high but still below the transition
temperature, one looks for thermal modifications of hadrons like a
modification of the mass, or a thermal width. Such a modification may
lead to, e.g., a relativistic Breit-Wigner form for the spectral function,
\beq
 \sigma_\had(p;T) \simeq Z \epsilon(p_0) \left(\frac2\pi\right)
   \frac{\Gamma m_\had}{(p^2-m_\had^2)^2 + \Gamma^2 m_\had^2},
\eeq{breit}
with both $\Gamma$ and $m$ possibly dependent on $T$.
A similar form of the spectral function is introduced at zero temperature
for a resonance. In that case this form is useful when the unitarity of
the S-matrix is not an issue. When it is, then a more careful analysis
is required where the decay products of the resonance are explicitly
included. The treatment of the scalar meson in \apxn{scl} is an example.

If we ignore such issues, then the screening correlator is connected to
the spectral function, as usual, through the integral \cite{laermann}
\beq
 S(z,T) = \int_{-\infty}^\infty \frac{dp_z}{2\pi} e^{ip_zz} \int_0^\infty 
   \frac{2dp_0}{p_0} \sigma(p_0,\vec0_\perp,p_z;T).
\eeq{scrn}
It is easy to see that a change in the spectral function from the stable 
form \eqn{stable} to a Breit-Wigner form \eqn{breit} causes the  
large-$z$ behavior of the screening correlation function to change to
\beq
   S(z,t) = 
    \frac12\left[ {\rm e}^{m_\had(N_L/2-z)} \cos\left(\frac{\Gamma z}4\right)
  + {\rm e}^{-m_\had(N_L/2-z)} \cos\left(\frac{\Gamma(N_L-z)}4\right) \right],
\eeq{bwcor}
assuming $\Gamma\ll m_\had$. One can see from this that the effect of
a width as small as $\Gamma/m \lesssim 0.1$, will be difficult to
detect when the error bars are as small as a few percent. However,
$\Gamma/m\simeq 0.2$ should be visible, provided the spectral function
of \eqn{breit} can be used.

On the other hand, the problem with the assumed Breit-Wigner form
becomes clear if one examines effective masses instead of correlation
functions. Then one finds that the effective masses would increase with
$z$, contrary to the behaviour seen for the PS, V and AV. The reason
is that in the lattice field theory, the unitarity of the S-matrix is
enforced through the formulation. As a result, one cannot, in general,
use a Breit-Wigner approximation to the spectral function in thermal QCD.

Next we make a detailed comparison of the behavior of the screening
correlator with the corresponding correlators in FFT.  We used wall
sources for FFT, and compared the effective masses, $m(z)$, obtained on
$32^2\times8\times N_z$ lattices (with $N_z=80$), with those obtained
in the interacting theory.

\bef[tbh]
\begin{center}
\incfig{0.6}{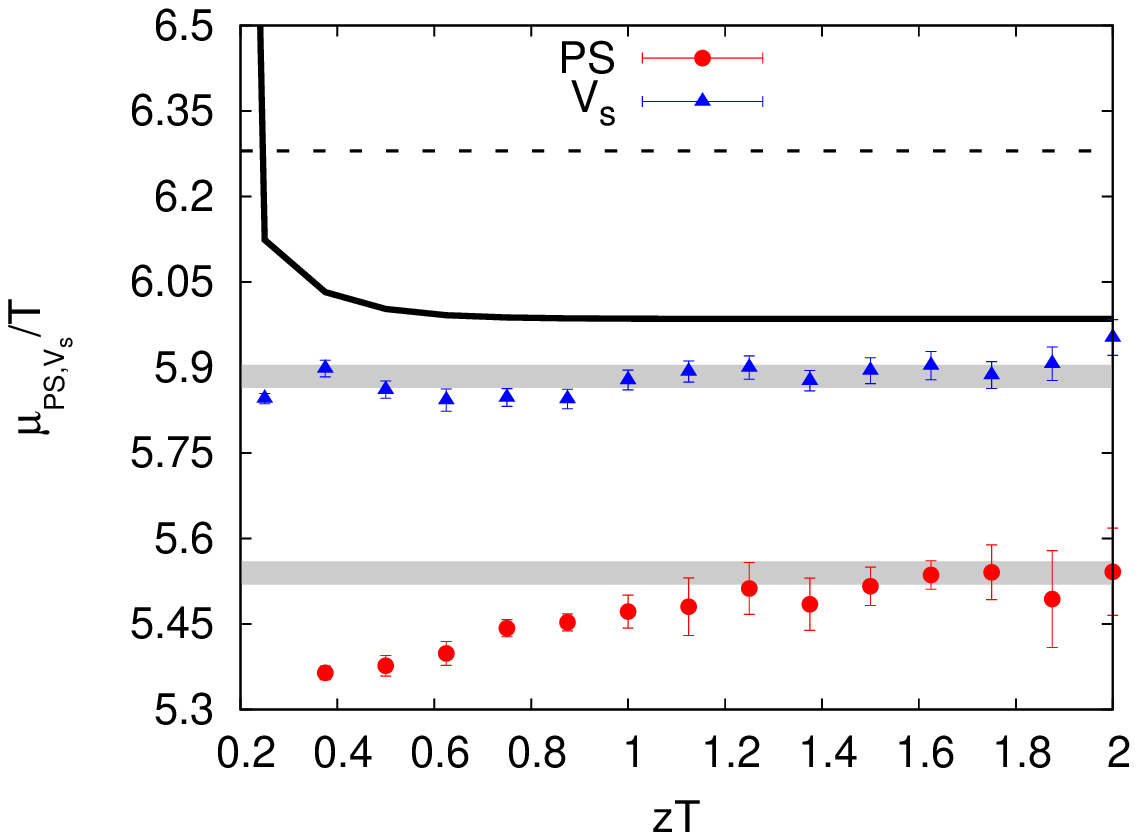}
\incfig{0.6}{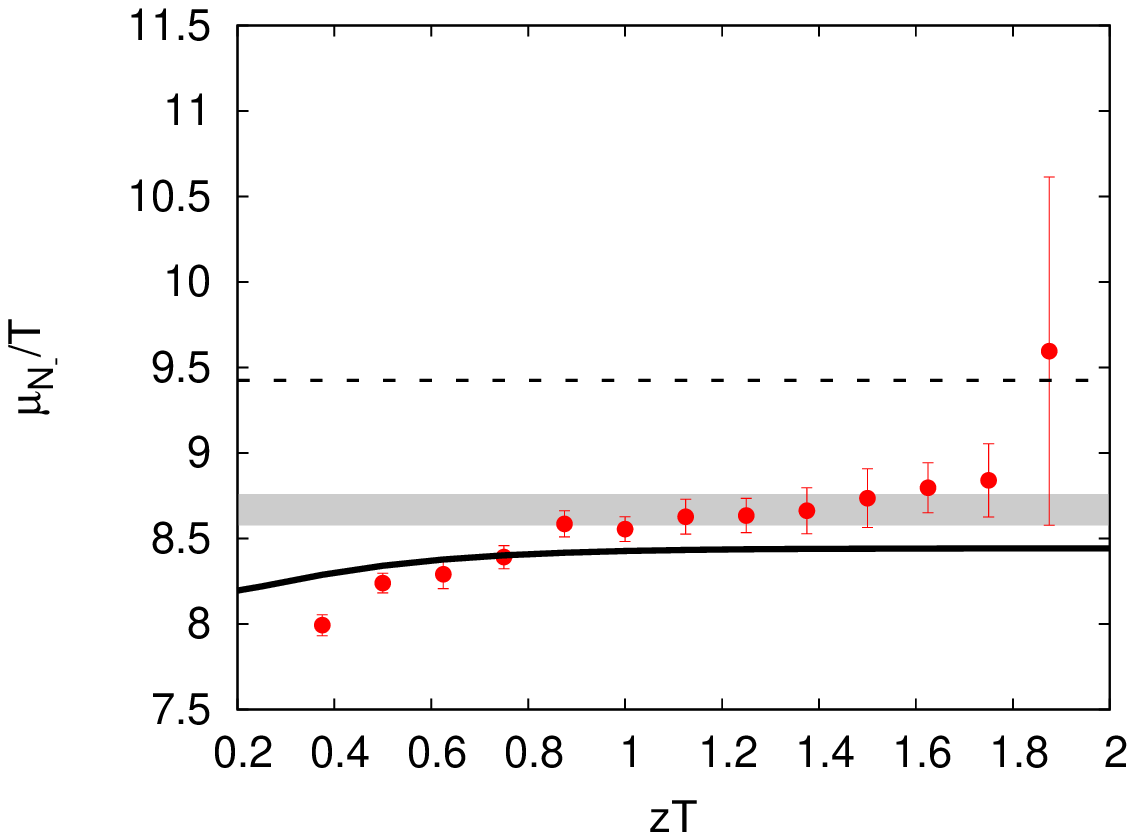}
\end{center}
\caption{Effective masses, in units of temperature, for meson and
nucleon sources, at $T=1.5T_c$ and in FFT. The black continuous curve
denotes the FFT screening masses for $N_t=8$. The black dashed line
shows the continuum value, $\mu=2\pi T$ \cite{FF01,mtc}.  The data points
and the bands show the effective masses and the fitted estimate for the
asymptotic value, respectively.}
\eef{FT}

\bef[tbh]
\begin{center}
\incfig{0.6}{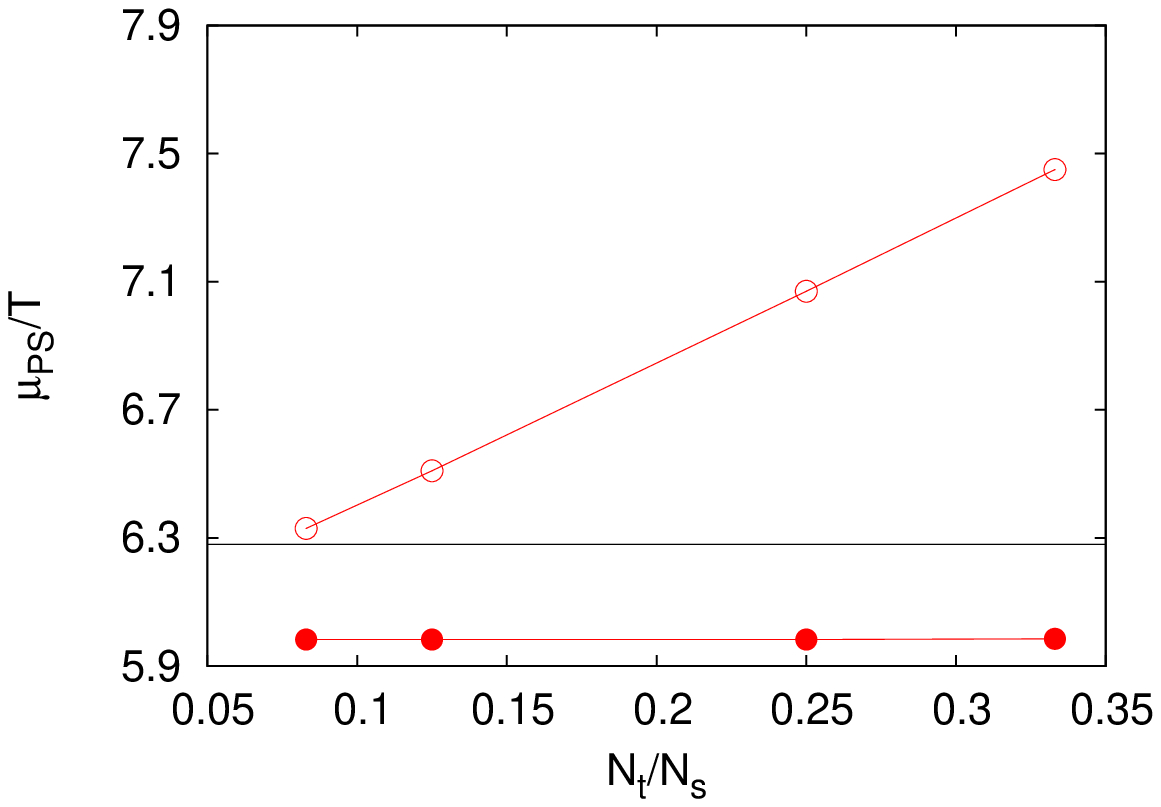}
\incfig{0.6}{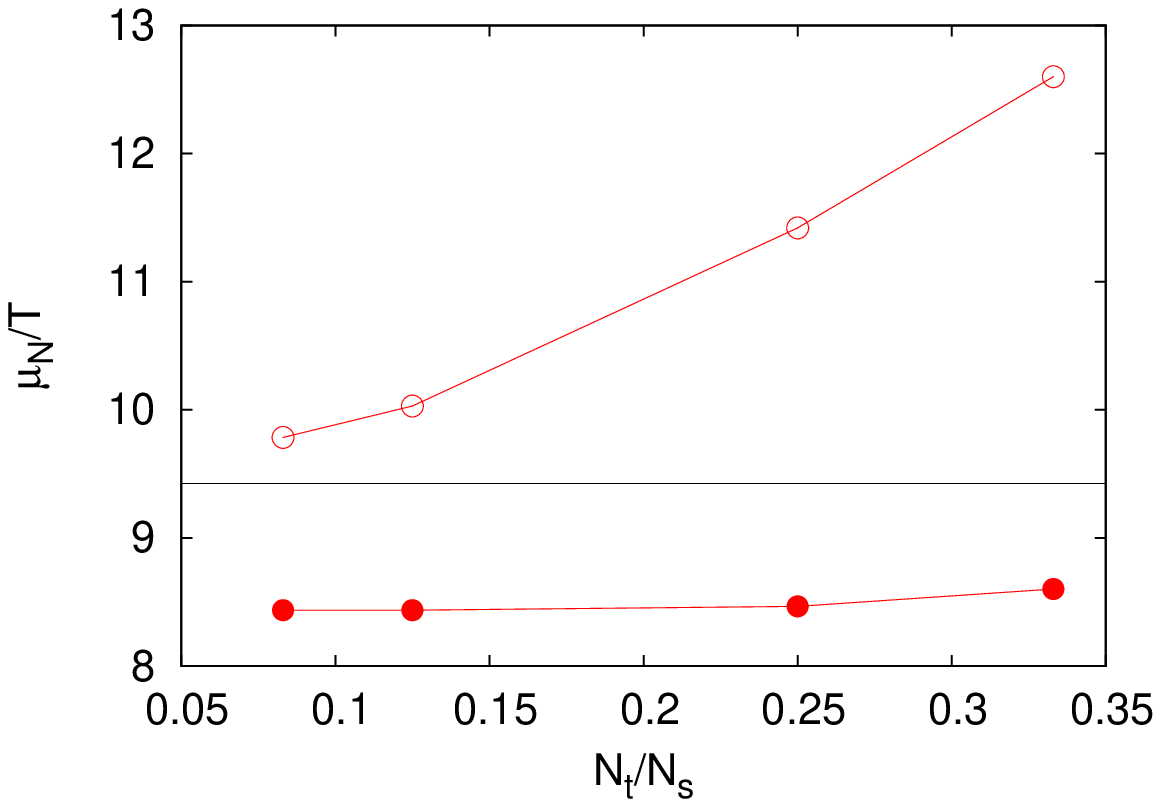}
\end{center}
\caption{Mesonic and nucleonic screening masses in FFT as a function of
$N_t/N_s=1/LT$ for $N_t = 8$. The filled symbols are estimates from wall
source quark propagators, while the unfilled symbols are estimated using
point sources. In the latter case, since there is no plateau in the
local mass, we use the prescription of \cite{Wis01} and use the local
mass at distance $N_z/4$.}
\eef{FSE}

In \fgn{FT} we compare the PS and V screening masses for our lightest
quarks (corresponding to $m_\pi/T=1.52$ at $1.5T_c$) to 
screening masses in FFT with similar bare quark mass.  In the free theory,
the screening correlator for wall source in these two channels are
equal. The results for the FFT were obtained on a lattice with
$N_z=80$ using the estimator $\ln C(z)/C(z+1)$ for the screening masses.
The plateaus in the effective masses shown is due to the use of a wall
source.  When a point source is used, neither the free theory nor the
interacting theory shows a plateau \cite{Wis01}. These checks give us
confidence that we are able to extract the asymptotic behavior of the
screening correlator.

The effective masses for the negative parity nucleon are also shown in
\fgn{FT}.  In the interacting theory, a reasonable plateau is obtained
in the effective mass, $C(z) \sim \exp (\mu(z) \times (N_z-z))$.
The plateau is more pronounced in our data for heavier quarks.  We see
that the effective mass in the interacting theory is larger than that
in the free theory.

Since we use $N_s/N_t=LT=4$ we also check finite volume effects in a
theory of free quarks.  In \fgn{FSE} we show the screening masses for
pseudoscalar meson and nucleon on lattices with $N_t=8$ and varying
$N_s$. With wall sources we find essentially no finite volume effect
in either channels already for $LT=4$.  Since finite volume effects are
larger in the free theory than in the interacting theory, therefore finite
volume effects in our studies are expected to be small.  In contrast,
point sources give large finite volume effects \cite{Wis01}.

\section{Scalar Meson\label{s:scl}}

\bef[tbh]
\centering
\subfigure[]{\incfig{0.1}{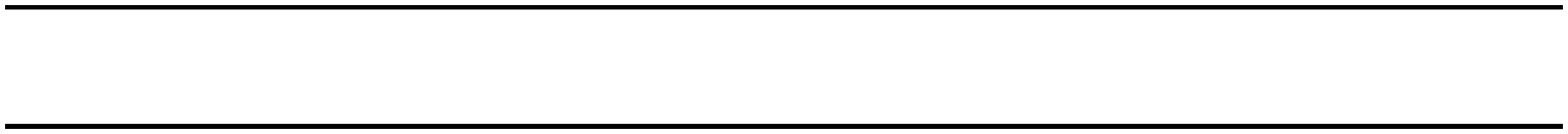}} \qquad
\subfigure[]{\incfig{0.1}{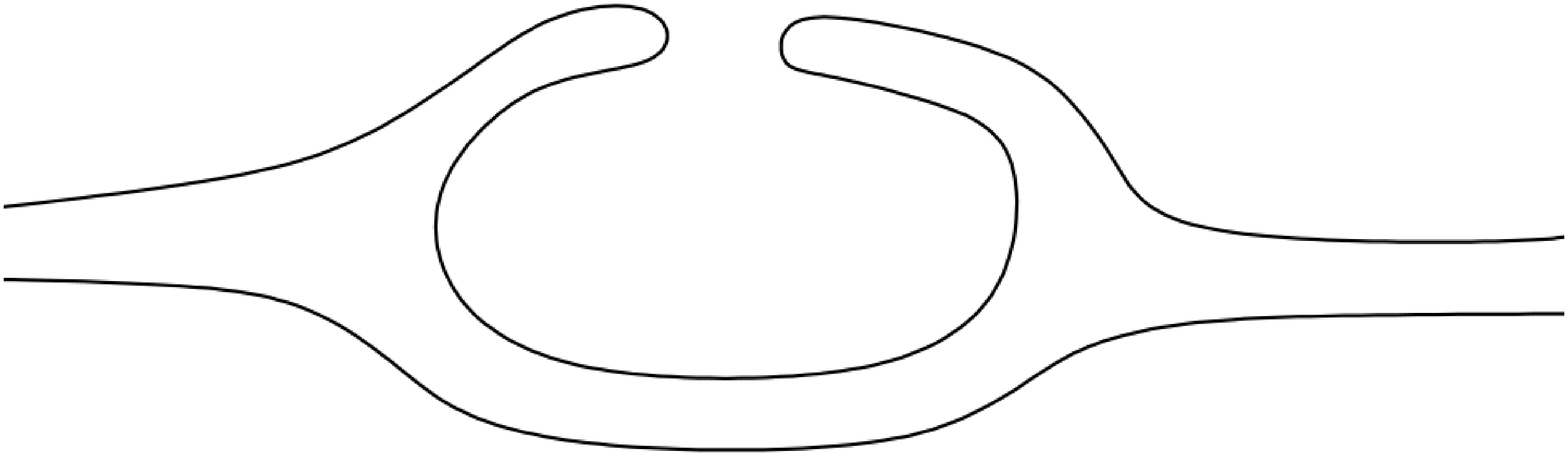}} \qquad
\subfigure[]{\incfig{0.1}{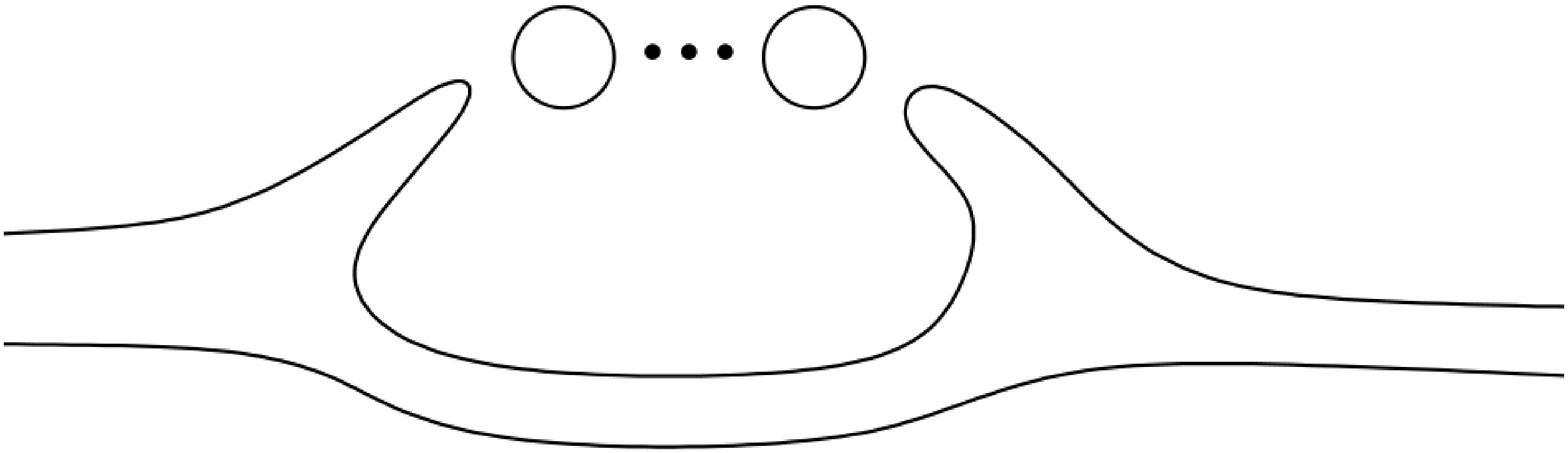}}
\caption{Quark line diagrams which couples with scalar propagator in the full 
theory. In the quenched theory (c) is absent.}
\eef{qualindiag}

The non-trivial behaviour of quenched scalar correlators may be
understood diagrammatically.  In full QCD the scalar propagator has
three kinds of topologies as shown in the figure above \cite{WB01}. The
third topology is absent in the quenched theory, as a result of which
the scalar correlation function becomes negative when the quarks are
light enough \cite{WB01,NM01,SP01}.  The missing diagram corresponds
(for an isovector scalar) to the propagation of two different, but mass
degenerate, pseudoscalars in the intermediate states; this is usually
called a $\pi\eta'$ (even in the two-flavour case). Due to the missing
diagram, the quenched theory lacks reflection positivity.

In full QCD one would be able to fit the scalar propagator with a simple
$\cosh$ term. However, in the quenched theory the contribution of the
missing term has to be cancelled (a quenched ghost) \cite{WB01}. This
leads to a complicated spectral function resulting in the fit form
which can be parametrized as \cite{WB01,NM01,SP01} 
\beqa 
   S^{Q\chi PT}(t) &=& b \left\{
  \exp{[-m_\SC t]} + \exp{[-m_\SC(N_t - t)]} \right\} \nonumber \\
   &&\quad -a_\SC \left\{ (1+m_{\pi}t)\exp({-E_{\eta'\pi}t}) +
 (1+m_{\pi}(N_t - t)) \exp({-E_{\eta'\pi}(N_t -  t)}) \right\} .
\eeqa{QChPT_long_t_beh_sc}
The negative term is explicitly due to the quenched ghost. $m_{\pi/\SC}$
are the masses of the PS and S meson, $E_{\eta'\pi}$ is the energy of the
ghost state ($E_{\eta'\pi}=2m_\pi-\eint$) and $-a_\SC$ is the coupling
to the ghost.  Since the ghost exactly cancels a physical term, these
two parameters are physical. For sufficiently small $m_\pi$ the second
term dominates at intermediate distances, leading to negative values
for the correlator.

In our analysis of the scalar propagator we take the long-distance part
of the propagator and extract from it both $a_\SC$ and $E_{\eta'\pi}$. Using
the renormalized propagator, we find the physical value of the coupling
$a_\SC$. These results are reported in \tbn{fitsc}.

The spectral function of the finite temperature scalar screening
correlator is not known. However, as for the other states, we will assume
that the form of the spectral function for $T<T_c$ is the same as at
$T=0$, but the parameters could depend on $T$. With this assumption we
will use the form in \eqn{QChPT_long_t_beh_sc} at finite temperature
below $T_c$.

\end{document}